\documentclass[%
12pts,
english,
showpacs,
twocolumn,
reprint,
epsf,
superscriptaddress,
amsmath,amssymb,
aps,
prb,
floatfix,
]{revtex4-2}

\pdfoutput=1
\usepackage[utf8]{inputenc}
\usepackage{hyperref}
\usepackage{amsmath}
\usepackage{marvosym}
\usepackage{amssymb,stmaryrd}   
\usepackage{bbm}
\usepackage{babel}
\usepackage{bm}
\usepackage{color}
\usepackage{natbib}
\usepackage{graphicx,subfigure}       
\usepackage{epstopdf}
\usepackage{float}
\usepackage{color,graphicx}
\graphicspath{ {images/} }
\usepackage{dcolumn}
\usepackage{bm}
\usepackage{wrapfig}
\usepackage{mathtools}
\usepackage{tikz}  
\usetikzlibrary{arrows,shapes,trees,positioning}  
\usepackage{comment} 
\usepackage{siunitx}

\usepackage{appendix}
\usepackage{hyperref}            
\usepackage{textcomp}
\hypersetup{
	colorlinks=true,
	linkcolor=blue,
	citecolor=blue,
	urlcolor=blue,
}
\usepackage{ragged2e}

\begin{document}


\title{Unusual magnetotransport and anomalous Hall effect in quasi-two-dimensional van der Waals ferromagnet Fe$_4$GeTe$_2$}

\author{Riju Pal}
\affiliation{Department of Condensed Matter Physics and Material Sciences, S. N. Bose National Centre for Basic Sciences, Block–JD, Sector–III, Salt Lake, Kolkata, 700106, India}

\author{Buddhadeb Pal}
\affiliation{Department of Condensed Matter Physics and Material Sciences, S. N. Bose National Centre for Basic Sciences, Block–JD, Sector–III, Salt Lake, Kolkata, 700106, India}

\author{Suchanda Mondal}
\affiliation{Saha Institute of Nuclear Physics, HBNI, 1/AF Bidhannagar, Calcutta 700064, India}

\author{Prabhat Mandal}
\affiliation{Department of Condensed Matter Physics and Material Sciences, S. N. Bose National Centre for Basic Sciences, Block–JD, Sector–III, Salt Lake, Kolkata, 700106, India}

\author{Atindra Nath Pal}
\affiliation{Department of Condensed Matter Physics and Material Sciences, S. N. Bose National Centre for Basic Sciences, Block–JD, Sector–III, Salt Lake, Kolkata, 700106, India}



\begin{abstract}

  Fe$_4$GeTe$_2$, an itinerant van der Waals (vdW) ferromagnet having Curie temperature (T$_C$) close to room temperature ($\sim 270$~K), attracted a lot of attention due to its unique magnetic properties. In particular, it exhibits another transition (T$_{SR}$ $\sim$ 120~K) where the easy axis of magnetization changes from in-plane to the out-of-plane direction in addition to the ferromagnetic transition. While the nature of magnetic properties in Fe$_4$GeTe$_2$ has been investigated to some extent, the role of various scattering mechanisms in electronic transport has not been investigated in detail. Here, we have studied the magnetotransport in a multilayer Hall bar device fabricated on 300~nm Si/SiO$_2$ substrate. Interestingly, the  zero field resistivity shows a negligible change in resistivity near the ferromagnetic (FM) transition unlike the typical metallic ferromagnet, whereas, it exhibits a dramatic fall below T$_{SR}$. Also, the resistivity shows a weak anomaly at T $ \sim $ 38~K (T$_Q$), below which the resistivity shows a quadratic temperature dependence according to the Fermi liquid behavior. Temperature-dependent Hall data exhibits important consequences. The ordinary Hall coefficient changes sign near T$_{SR}$ indicating the change in majority carriers. In a similar manner, the magnetoresistance (MR) data also shows nonmonotonic behavior with a significantly large negative MR ($\sim 11\%$ at 9~T) near T$_{SR}$ and becomes positive below T$_Q$. As positive MR signifies the dominance of the orbital effect, negative MR reflects the reduction of electron-magnon scattering by the application of an external magnetic field. The observations of anomaly in the resistivity, sign-change of the ordinary Hall coefficient and maximum negative magnetoresistance near T$_{SR}$, together suggest a possible Fermi surface reconstruction associated with the spin reorientation transition. Furthermore, analysis of the Hall data reveals a significant anomalous Hall conductivity from $\sim 123~\Omega^{-1}$ cm$^{-1}$ (at T $\approx$ 5~K) to the maximum value of $\sim 366~\Omega^{-1}$ cm$^{-1}$ near T$_{SR}$. While the low-temperature part may originate due to the intrinsic KL mechanism, our analysis indicates that the temperature-dependent anomalous Hall conductivity (AHC) is primarily appearing due to the side-jump mechanism as a result of the spin-flip electron-magnon scattering. Our study demonstrates an interplay between magnetism and band topology and its consequence on electron transport in Fe$_4$GeTe$_2$, important for its future application in spintronic devices.  

\end{abstract}

\maketitle


\section{\label{sec:intro}Introduction}

The discovery of quasi-two-dimensional van der Waals (vdW) magnets \cite{Huang2017a, Gong2017a, Song2018a, Klein2018, Burch2018, McGuire2015} has opened up a new platform for investigating low-dimensional magnetism and its possible application in two-dimensional (2D) spintronic devices \cite{Zhong2017, Samarth2017, Tan2018, Kim2021a}. With the recent developments, the family of iron-based vdW magnets \cite{Zhang2019, Kang2020, Wu2021, Meng2021}, like Fe$_n$GeTe$_2$ (n = 3, 4, 5) (FnGT) \cite{Fei2018a, Deng2018, Kim2018e, Seo2020b, Mondal2021, Bera2023, Bera2022b, Fujita2022a, Yamagami2022, Ohta2021, Tan2021, Alahmed2021, Deng2022}, especially Fe$_4$GeTe$_2$ (F4GT) \cite{Seo2020b, Mondal2021, Bera2023} and Fe$_{5-x}$GeTe$_2$ (F5GT) \cite{May2019c, Zhang2020a, Liu2022a}, have attracted immediate attention due to their ferromagnetic transition temperature (T$_C$) closed to room temperature. As the Mermin-Wagner theorem \cite{Mermin1966} in the 2D limit dictates that there is no spontaneous magnetic order at finite temperature, the uniaxial magnetocrystalline anisotropy stabilizes the long-range order in these vdW systems against the thermal fluctuations. The enhanced T$_C$ is achieved by increasing the exchange interaction as a result of the metal-rich unit cell \cite{Bander1988a, Hall1991, Seo2020b}. However, the magnetism in these materials is rather complex in nature as compared to a typical ferromagnet, due to the presence of different inequivalent Fe atoms in the unit cell. For example, Fe$_3$GeTe$_2$ (F3GT) possesses antiferromagnetic order and noncollinear spin structure below 152~K \cite{Yi2017b}, also an unusual magnetic behavior was observed in F5GT at low temperature due to structural ordering of one of the Fe-atoms in the unit cell at $\sim$ 120~K \cite{May2019c}. More recently, it has been reported that F4GT exhibits a change in easy axis of magnetization when cooled below $\sim$ 110~K termed as the 'spin reorientation transition' (SRT) \cite{Seo2020b, Mondal2021}, making it magnetically quite different from the other two family members. A similar spin reorientation transition was observed in materials like Fe$_3$Sn$_2$ \cite{Wang2016e, Kumar2019a, Biswas2020, Fenner2009a}, Nd$_2$Fe$_{14}$B \cite{Xiao2020}, TbMn$_6$Sn$_6$ \cite{Yin2020a, Zajkov2000, Malaman1999, Clatterbuck1999}, LiMn$_6$Sn$_6$~\cite{Wang2023}, NdCrSb$_3$ \cite{Chen2023}, La$_{0.4}$Sm$_{0.3}$Sr$_{0.3}$MnO$_3$ \cite{Aparnadevi2013} etc. The interplay between exchange and anisotropy is possibly the reason for this spin reorientation \cite{Seo2020b}. The recent transport measurements indicate that the SRT may lead to Lifshitz transition in the electronic structure which may lead to unusual magnetotransport and anomalous Hall effect \cite{Wang2022a, Wang2023}. While in the case of F4GT, an anomaly in the specific heat was seen, indicating that it is indeed a thermodynamic phase transition \cite{Bera2023}, its consequence on electron transport is still elusive.

\begin{figure*}[ht!]
	\centering
	\includegraphics[width=18cm]{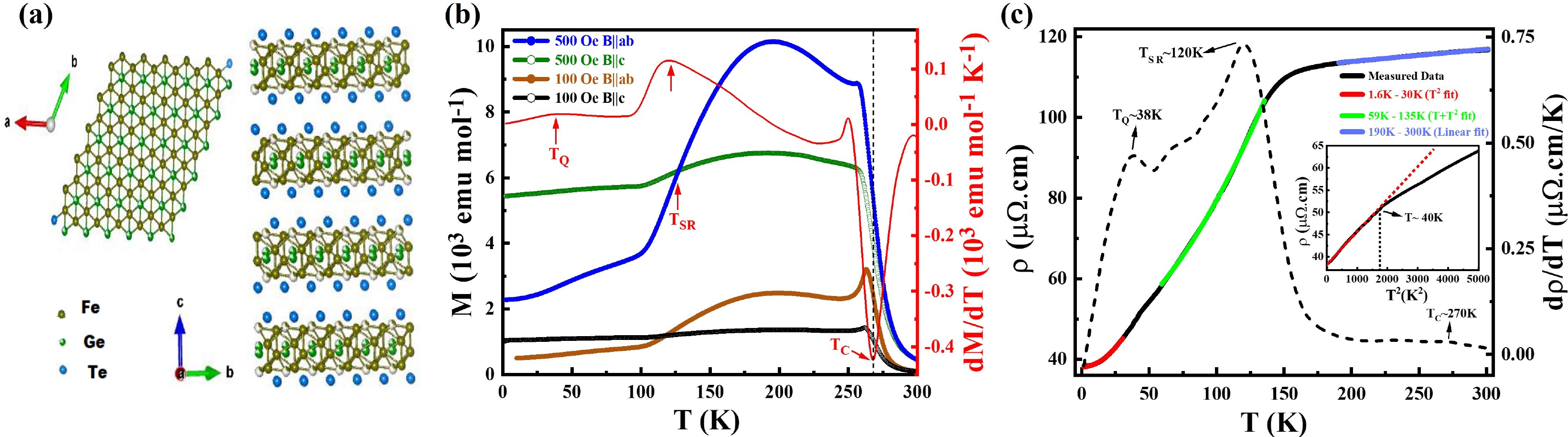}
	\caption{(a) Schematic diagram of the structure of F4GT crystal. ABC stacking of monolayer F4GT is shown in the right-handed picture. (b) The dc magnetization data as a function of temperature is plotted with applied field 500 Oe and 100 Oe along B $ \parallel $ ab and B $ \parallel $ c of the crystallographic axis of F4GT crystal, measured in ZFC condition. The red curve shows the temperature derivative of the magnetization (dM/dT) data of 500 Oe with B $ \parallel $ c, indicates three transitions, Curie temperature ($ T_{C} $)  around 270~K, the spin-reorientation transition (T$ _{SR} $) at around 120~K, and a small kink ($ T_{Q} $) at around 38~K. (c) Temperature dependence zero field electrical resistivity curve (black color) and its temperature derivative (black dotted) is plotted with current through in-plane direction (I $ \parallel $ ab plane). Theoretic fits to the temperature dependence of zero-field resistivity data at different temperature ranges is indicated with red, green, and light blue color. Inset: $ \rho $ vs $ T^{2} $ shows a clear change in the slope above 38~K.}
	\label{fig:Fig1}
\end{figure*}

Besides the unusual magnetic properties, electronically these materials possess interesting features. All the members of the FnGT family are predicted to be semimetal as the DFT-based calculations show multiple band crossing at the Fermi level \cite{Sau2022}. More importantly, the presence of different crystal symmetry and the spin-orbit coupling (SOC) may suggest a topologically nontrivial phase with unusual effects induced by the chiral anomaly like negative magnetoresistance or nonlinear conductivity in the diffusive limit \cite{Kim2018e}. Furthermore, the broken time-reversal symmetry (TRS) in these topological phases hints a more exotic ground state leading to observations like large intrinsic anomalous Hall effect (in F3GT) \cite{Wang2017} or unusual magnetotransport behavior at low temperatures \cite{Ke2020b}. With increasing temperature, these systems often exhibit nonmonotonic transport features connected to temperature-dependent magnetic behavior \cite{Jones2022}. While the low-temperature transport is relatively easy to address as the effect of inelastic electron-phonon or electron-magnon interactions is negligible compared to the intrinsic effect, the role of these competing interactions on the transport behavior at intermediate or high temperatures is not fully established \cite{Saha2023}. 

\begin{figure*}[t!]
	\centering
	\includegraphics[width=18cm]{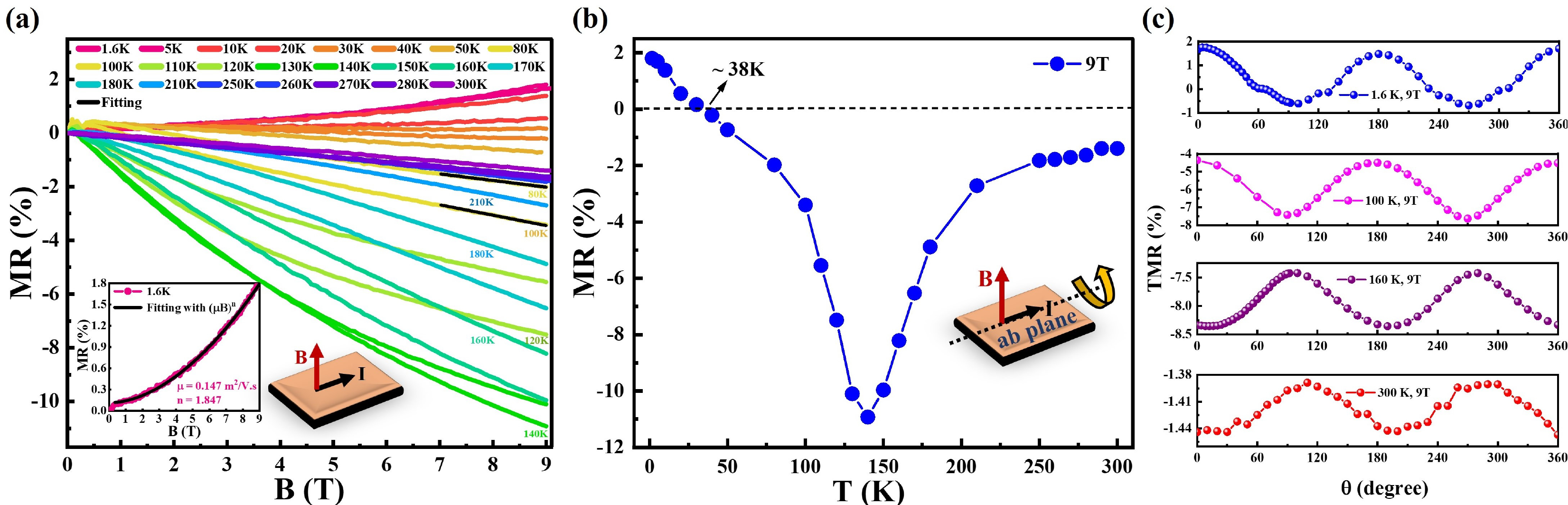}
	\caption{(a) Magnetic field dependence of magneto-resistance (MR) measured from temperature 1.6~K upto 300~K. Here, the external magnetic field is applied parallel to the c-axis and the current is in the ab plane of the F4GT sample. Black lines are the fitting of Eq. \ref{Eq.4} at 80~K and 100~K. Inset: Low-temperature MR is fitted with Eq. \ref{Eq.3}. (b) Temperature dependence of MR is plotted with applied field 9~T along the c-axis of the crystal, indicating a maximum negative MR at around T$_{SR}$. (c) Angle-dependent transverse magneto-resistance (TMR) at 9 T at different temperatures shows a definite spin reorientation between 100~K and 160~K and it promotes the anisotropic behavior of the crystal.}
	\label{fig:Fig2}
\end{figure*}

Here, we study in detail the temperature-dependent electronic and magnetotransport behavior of F4GT. A single layer of F4GT consists of seven atoms (Fe$_1$ and Fe$_2$ arranged on both sides of the Ge atomic plane and they are connected with Te atoms directly on both sides) as shown in Fig. \ref{fig:Fig1}(a). The stacking of these F4GT monolayers forms a rhombohedral structure with space group $ R\bar{3}m $ \cite{Seo2020b}. Electronically this has been predicted to be a different class with nontrivial topology, unlike F3GT or F5GT. Single crystal of F4GT was prepared by the chemical vapor transport (CVT) methods exhibiting the ferromagnetic T$_C\sim 270$~K, followed by the SRT at $\sim$ 120~K (see Fig. \ref{fig:Fig1}(b)). By fabricating multilayer Hall bar devices on predefined Ti/Au contacts using the dry transfer method (see Supplementary Information S1 for details), we studied the temperature-dependent resistivity, magnetoresistance, and Hall effect from room temperature (300~K) down to 1.6~K. In particular, we observe the direct consequences of the SRT, leading to the change in carrier types (electrons to holes) through a Lifshitz transition, confirmed by the temperature-dependent ordinary Hall coefficient. Similarly, a strong decrease in resistivity followed by the enhanced negative MR and the maximum anomalous Hall conductivity were also observed near the SRT. While the analysis of the resistivity data uncovers the role of different inelastic scattering mechanisms like electron-electron, electron-magnon, and electron-phonon interaction in the different temperature ranges, we find that electron-magnon is also responsible for the temperature-dependent MR or anomalous Hall effect. Finally, we report a new electronic transition (T$_Q$) near $\sim$ 38~K, below which the resistivity behaves like a Fermi-metal with quadratic temperature dependence, along with a weak positive MR.

\section{Results and Discussions} \label{sec:results}

\subsection{Resistivity} \label{sec:resistivity}

Figure~\ref{fig:Fig1}(c) shows the temperature-dependent in-plane resistivity ($ \rho_{xx} $), measured with a constant ac excitation of 50 $\mu$A at zero magnetic field. This exhibits a metallic behavior, with almost negligible change near the FM transition. However, the resistivity falls dramatically near SRT. $ \rho_{xx} $ shows a weak anomaly at $T_{Q}$ ($\sim $ 38~K), indicates a clear kink in the $ d\rho_{xx}/dT $ curve (black dotted line in Fig.~\ref{fig:Fig1}(c)), whose consequence on the transport will be discussed. The residual resistivity ratio (RRR = $\rho_{xx}$(300~K)/$\rho_{xx}$(1.6~K)) value of the exfoliated F4GT device is 3.04, which is consistent with the previous report \cite{Seo2020b}.  The conductivity ($ \sigma $) of our sample at $ T_{C}$ ($ \sim $ 270~K) is $ \sim 8.6 \times 10^{5}~\Omega^{-1}$m$^{-1} $, indicating relatively higher conductivity compared to the other 2D ferromagnets \cite{McGuire2015, Chen2013, Kim2018e} and consistent with the reported value in this material \cite{Seo2020b}. 

According to Matthiessen's rule, the total resistivity of a metallic ferromagnet consists of all the resistivity contributions coming from various scattering mechanisms and they are additive within each conduction band \cite{Raquet2002, Jena2020}. The temperature dependence of longitudinal resistivity can be written as:
\begin{equation}
	\rho_{xx} (T) = \rho_{0} + \rho_{e-p}(T) + \rho_{e-e}(T) + \rho_{e-m}(T,B)
	\label{Eq.1}
\end{equation}
where $ \rho_{0} $ is the residual resistivity arising due to the temperature-independent elastic scattering from the static defects. $ \rho_{e-p} $, $\rho_{e-e}$ and $\rho_{e-m}$ are the inelastic electron-phonon, electron-electron and electron-magnon  scattering contributions, respectively. Among these, $\rho_{e-p}$ varies linearly with temperature ($ \propto T $) and both $ \rho_{e-e}$ and $ \rho_{e-m}$ exhibit quadratic behavior with temperature ($ \propto T^{2} $). As the electron-magnon term is strongly dependent on the magnetic field, whereas, other terms remain insensitive, the field-dependent resistivity can be used to identify the actual mechanism.

The temperature-dependent resistivity data in Fig. \ref{fig:Fig1}(c) shows different natures in different temperature regimes. At very low-temperature below 30~K, $ \rho $ follows a perfectly quadratic behavior ($ \rho \propto T^{2} $), corresponding to either electron-electron (e-e) scattering or electron-magnon (e-m) scattering term. To identify the actual mechanism, we measured the temperature-dependent resistivity at a high magnetic field (9~T) in both the in-plane and out-of-plane direction of the crystal (See Supplementary Information, section S3). It is observed that the temperature-dependent resistivity is almost independent of the magnetic field, which suggests that the dominant scattering mechanism is indeed the electron-electron interaction, confirming Fermi-liquid behavior. The magnitude of the coefficient of the quadratic term is the measure of the electron-electron scattering rate. In our case, we found the value of this coefficient is 7.73 $\times 10^{-9}$ $ \Omega $ cm K$^{-2} $, which is nearly two orders of magnitude larger than the elemental ferromagnets like Fe, Co, and Ni, but comparable to the value of the semi-metals like Bi, graphite, etc. \cite{Behnia2022, Klein1996, Lin2015, Pariari2021, Wang2020l, Susmita2018}.  In the intermediate range (59~K - 135~K), the data can be fitted with the admixture of both linear and quadratic contributions (i,e. $\rho \propto (T+T^{2}) $). While the linear dependence corresponds to the electron-phonon coupling, the T$^2$ dependence signifies the electron-magnon scattering. This is evident from the fact that the resistivity has a strong dependence on the magnetic field in this regime and the coefficient of the T$^2$ term is reduced significantly at a high magnetic field (see Supplementary Information S3). At the high-temperature range (T $>$ 190~K), well above the spin-reorientation transition, a complete linear dependence of resistivity with temperature is observed ($ \rho \propto T $), indicating the dominance of the electron-phonon scattering mechanism. 

\subsection{Magnetoresistance} \label{sec:MR}
We next concentrate on the temperature-dependent magneto-resistance (MR) data. Fig. \ref{fig:Fig2}(a) shows the MR ( = $((\rho_{xx}(B)-\rho_{xx}(0))/\rho_{xx}(0)) \times 100\% $) at different temperatures starting from 300~K down to 1.6~K with the applied magnetic field along the out-of-plane direction (c-axis) of the F4GT crystal. In order to eliminate the contribution from the Hall resistance due to a small misalignment of the electrodes, we symmetrize the data using the formula $ \rho_{xx}(B) = (\rho_{xx}(+B) + \rho_{xx}(-B))/2$. The high field (9 T) MR data (Fig. \ref{fig:Fig2}(b)) exhibit several interesting features. MR is predominantly negative in the range between $\sim 40$~K to 300~K and maximum near the SRT ($\sim 11\%$). At temperature, below T$_Q$, the MR becomes positive ($\sim$ 1.8\% at 1.6~K), unusual for a metallic ferromagnet. To determine the anisotropy of the crystal, we have performed the angle-dependent transverse magnetoresistance (TMR) measurements at 9 T at different temperatures, as in Fig. \ref{fig:Fig2}(c). While the low-temperature data below 100~K show cosine-like behavior, above the SRT (160~K and 300~K) it shows a phase shift by 90 degrees, depicting the easy axis change from the c-axis to the ab-plane with increasing temperature. (See Supplementary Information Fig. S4.2).

Typically the resistivity of a ferromagnetic material can be represented as a function of the electronic relaxation time ($ \tau $) and its field dependence as \cite{Taylor1968a, Porter2014}, 
\begin{equation}
	\rho_{total} = k_{1}(\omega_{c}\tau)^{n} + k_{2}(1/\tau)
	\label{Eq.2}
\end{equation}
where, $ \omega_{c} $ is the cyclotron frequency of the free carriers. Here, the first term is the classical contribution to the resistivity arising due to the constrained orbital motion of the free carriers under the Lorentz force and responsible for positive MR. The second term describes  the contributions from the various scattering mechanisms as explained in Eq. (\ref{Eq.1}). The field dependence of the orbital term can be expressed as, 
\begin{equation}
	\rho_{orb} \propto (\omega_{c}\tau)^{n} = (\mu B)^{n}
	\label{Eq.3}
\end{equation}
where $\mu $ is the mobility of the free carriers. Ideally, for a nonmagnetic metal, the exponent n = 2 \cite{Pippard1989MagnetoresistanceIM}, but for other systems like doped semiconductors \cite{Khosla1970}, ferromagnetic metallic thin films \cite{Raquet2002}, several spin-glass systems \cite{Yosida1957}, etc., it deviates from 2 and lies in between 1 and 2 ($1 < n < 2$). By fitting the positive MR at low temperature, we observe that the exponent, n, becomes close to 2 (See Fig. \ref{fig:Fig2}(a) inset and Supplementary information section S4.1 for fitting), confirming that the material behaves like a nonmagnetic metal below T$_Q$, consistent with the Fermi liquid behavior. 

At higher temperatures above T$_Q$, we have observed a crossover in MR from positive to negative (See Fig. \ref{fig:Fig2}(a)) and becomes maximum near the SRT ($\sim -11\%$ at 9 T). The enhanced MR is possibly associated with the dominance of electron-magnon scattering in this intermediate range, along with the possible change in electronic and magnetic properties associated with the spin reorientation transition. Similar behavior was observed in F5GT, where a structural ordering was observed through neutron scattering measurement near ($\sim 100$~K), which influences the electronic structure and magnetic moment of the system \cite{May2019c}. 
 
The role of electron-magnon scattering on negative MR was discussed in Ref.~\cite{Raquet2002} and an analytical expression was given, which is valid for magnetic field below 100 T and in the temperature range of $ T_{C} $/5 to $ T_{C} $/2, 
\begin{equation}
	\Delta\rho_{xx}(T,B) \propto \dfrac{BT}{D(T)^{2}}ln(\dfrac{\mu_{B}B}{k_{B}T})
	\label{Eq.4}
\end{equation}
where, D(T) is the magnon stiffness or magnon mass renormalization, $ \mu_{B} $ is the Bohr magneton and $ k_{B} $ is the Boltzmann constant. The high field MR data (B $>$ 7 T) above 50~K in the temperature range 80~K and 100~K ($ T_{C} /5 < T <  T_{C} $/2) can be well-fitted with Eq. (\ref{Eq.4}) (Fig. \ref{fig:Fig2}(a)), confirming that the non-saturated negative MR is originated primarily due to the suppression of electron-magnon scattering under the external magnetic field.

\subsection{Hall Measurements} \label{sec:Hall}

\begin{figure*}[ht!]
	\centering
	\includegraphics[width=14cm]{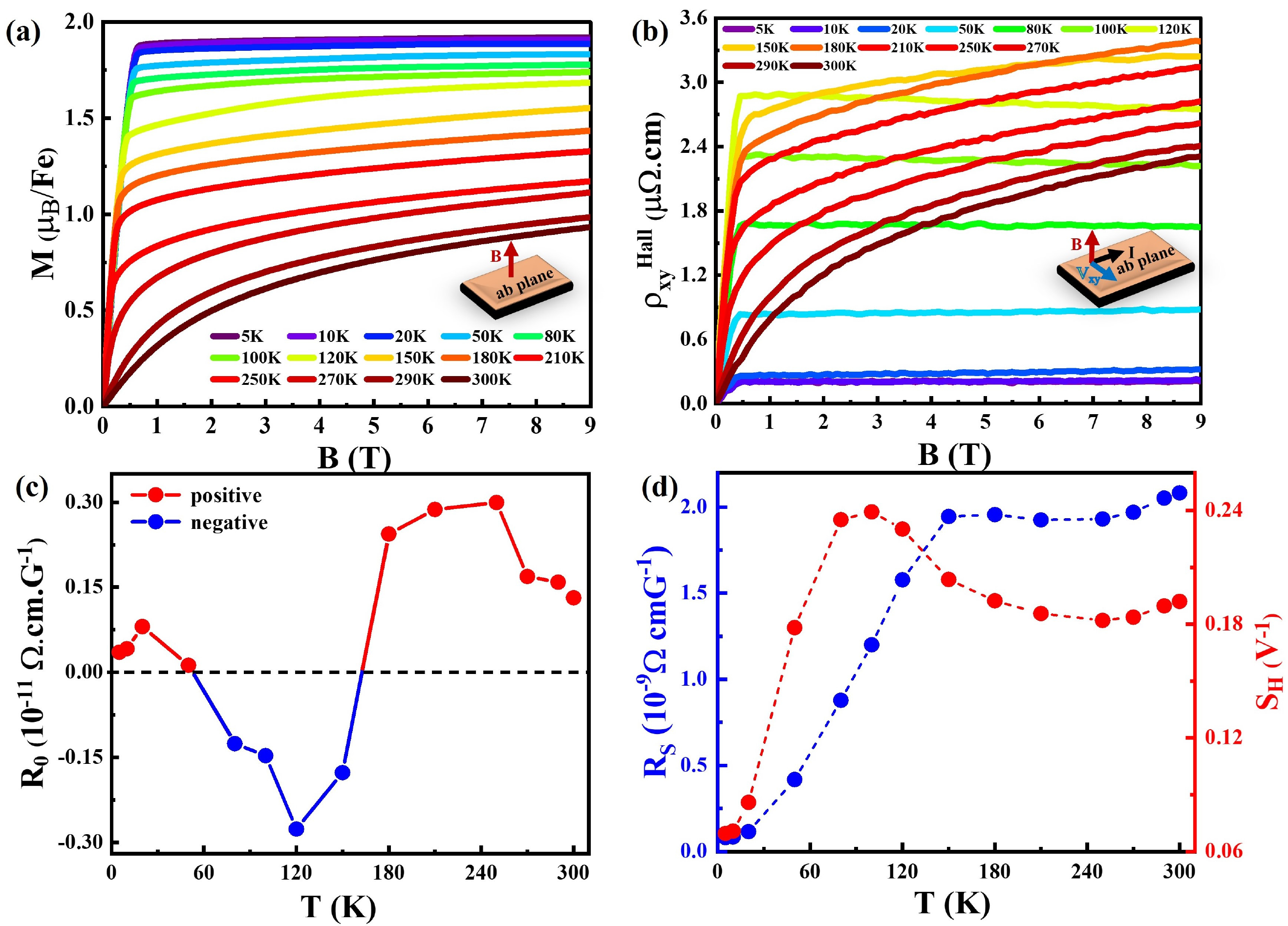}
	\caption{(a) Magnetic field dependence of magnetization (M) at different temperatures with B normal to the ab-plane of the crystal. (b) Magnetic field dependence of Hall resistivity ($ \rho _{xy} $) measured at temperatures from 5~K to 250~K, with current in ab-plane and magnetic field along the c-axis of the F4GT crystal. (c) Temperature dependence of ordinary Hall coefficient (R$ _{0} $) indicates both positive and negative values at different temperatures. (d) The temperature dependence of anomalous Hall coefficient $ R_{S} $ and scaling coefficient $ S_{H} = R_{S}/\rho_{xx}^{2} = -\sigma_{xy}^{AHE}/M $.}
	\label{fig:Fig3_OK}
\end{figure*}

Now we will focus on the Hall effect in F4GT. Typically, the transverse Hall resistivity ($\rho_{xy}$) of a ferromagnetic material can be described by an empirical formula \cite{Pugh1930, Pugh1932},
\begin{equation}
	\rho_{xy} = \rho_{xy}^{OHE} + \rho_{xy}^{AHE} = R_{0}B + \mu_{0}R_{S}M
    \label{Eq.5}
\end{equation}

Here, the first term is the ordinary Hall resistivity ($ \rho_{xy}^{OHE} $) and the second term represents the anomalous Hall resistivity ($ \rho_{xy}^{AHE} $). From the magnitude of the ordinary Hall coefficient, $ R_{0} $, one can calculate the effective carrier concentration (n) when a single band picture is valid and its sign determines the type of majority carriers present in the material. The anomalous Hall part is proportional to the spontaneous magnetization (M) and the proportionality constant, $ R_{S} $, is defined as the anomalous Hall coefficient.  In most ferromagnets, the magnetization (M) saturates above some critical field below the Curie temperature and $\rho_{xy}$ varies linearly with B. From the linear fitting of $\rho_{xy}$ in the high field regime, one can obtain $ R_{0}$ from the slope and the anomalous Hall resistivity ($ \rho_{xy}^{AHE} $) from the intercept.  Also, the anomalous Hall coefficient, R$_S$ can easily be calculated by using the relation, $\rho_{xy}^{AHE} = \mu_{0}R_{S}M_{S}$, where $ M_{S} $ can be extracted from the M-B curves Fig. \ref{fig:Fig3_OK}(a) above B $\geq$ 7 T.

However, the above-mentioned technique is not applicable when the magnetization does not saturate at a high field as can be seen in the temperature-dependent M-H data for F4GT (see Fig. \ref{fig:Fig3_OK}(a) and Supplementary Information S2). We  observe that the magnetization does not show complete saturation above T$_{SR}$ till B = 9 T (Fig. \ref{fig:Fig3_OK}(a)). To circumvent the problem, we incorporate the field dependence of magnetization in Eq. \ref{Eq.5} and use the modified equation for fitting as, 
\begin{equation}
	\frac{\rho_{xy}}{B} = R_{0} + \mu_0 R_{S}\frac{M}{B}
    \label{Eq.6}
\end{equation}
From that fitting of $ \frac{\rho_{xy}}{B} $ vs $ \frac{M}{B} $, we can easily determine the slopes and y-axis intercepts, which provide us the temperature-dependent values of $ R_{S} $ and $ R_{0} $, respectively. Both methods qualitatively provide equivalent results except some quantitative differences in the high-temperature range. Details of calculation procedures are explained in Supplementary Information section S5.

Figure \ref{fig:Fig3_OK}(b) demonstrates the in-plane Hall resistivity ($\rho_{xy}$) measured at different temperatures down to 1.6 K with an ac excitation current of 5~$\mu$A with the magnetic field along c-direction. In order to eliminate the longitudinal component due to small misalignment of the Hall electrodes, the transverse Hall resistivity ($\rho_{xy}$) was measured by taking the data for both positive and negative magnetic fields and then taking the anti-symmetric components of the transverse Hall resistivity by using the formula: $ \rho_{xy}(B)$ = $(\rho_{xy}(+B) - \rho_{xy}(-B))/2 $. The symmetrized transverse Hall resistivity is plotted as a function of magnetic field at different temperatures as in Fig. \ref{fig:Fig3_OK}(b). Here, the similarity in the nature of magnetic field dependence curves of transverse Hall resistivity and magnetization in the low-field region indicates the presence of anomalous Hall effect behavior.

\begin{figure*}[htbp!]
	\centering
	\includegraphics[width=14cm]{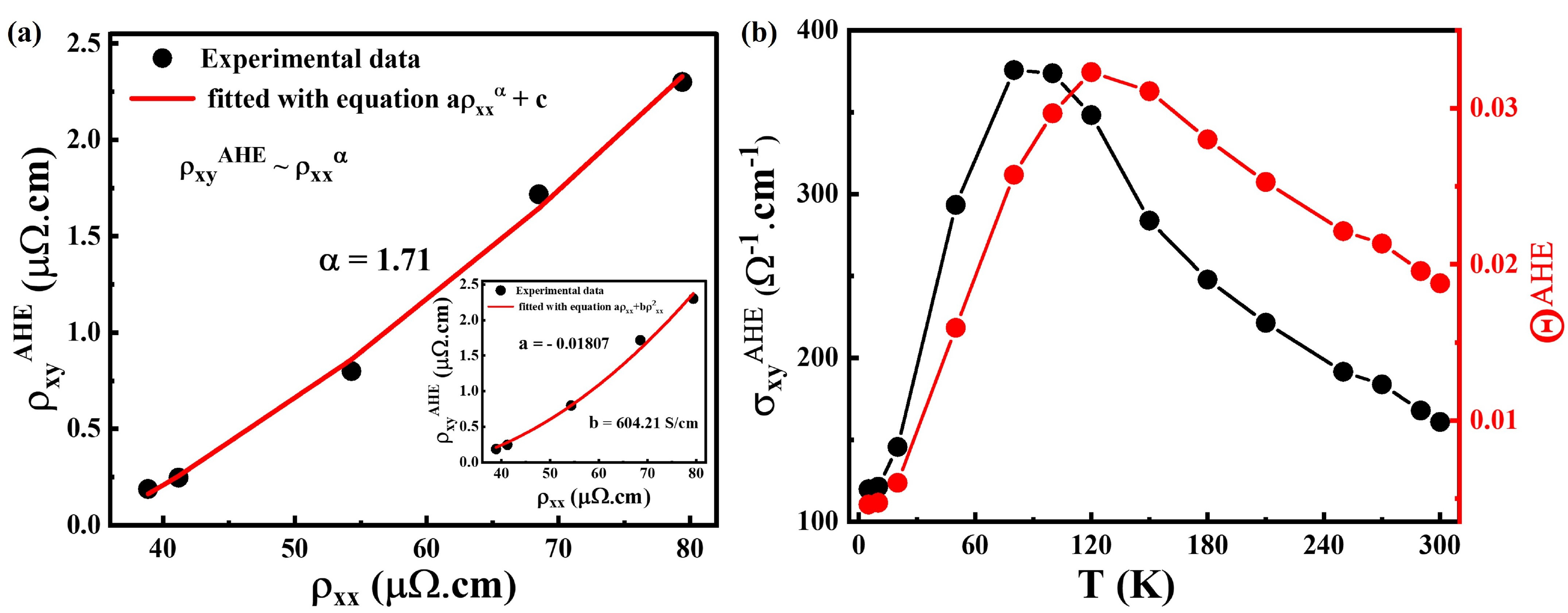}
	\caption{(a) Scaling behaviour of anomalous Hall resistivity ($ \rho_{xy}^{AHE} $) vs in-plane resistivity ($ \rho_{xx} $). The red curve indicates the fitting of the experimental data, which helps to understand the origin of the anomalous Hall effect. Inset; The data is fitted with the equation $ \rho_{xy}^{AHE} = a\rho_{xx}+ b\rho_{xx}^2 $ (b) Temperature dependence of anomalous Hall conductivity ($ \sigma_{xy}^{AHE} $) provides a large anomalous Hall conductivity of 365.8 $ \Omega^{-1}$cm$^{-1}  $ at 80 K. Temperature dependence of anomalous Hall angle $ \theta^{AHE} = \sigma_{xy}^{AHE}/\sigma_{xx} $ has an anomaly near the spin reorientation transition ($ T_{SR} $)}
	\label{fig:Fig4_OK}
\end{figure*}

First, we concentrate on the temperature-dependent behavior of the ordinary Hall coefficient, R$_0$, which shows a rather complex nature. Fig. \ref{fig:Fig3_OK}(c) shows that $R_{0}$ changes sign from positive (red) to negative (blue) near the SRT region ($\approx$ 130~K) and again becomes positive $\approx$ 50~K. This indicates the change in the majority carrier from electrons to holes, possibly due to the Fermi surface reconstruction or Lifshitz-like transition occurring as a result of the SRT \cite{Wang2022a, Wang2023}. The origin of this change of holes to electrons near $\approx$ 50~K is not clear, albeit, the dominance of pure electron-electron interaction along with the occurrence of positive MR below T$_Q$ indicate an electronic transition to a typical Fermi metal. We can compare this observed data of F4GT with the reported result \cite{Wang2017} of F3GT. Similar to F4GT, the Hall resistivity in F3GT is linear in field above a critical value. Unlike F4GT, the slope of the Hall resistivity ($R_{0}$) in F3GT is positive from 2 K to 300 K, indicating the hole-dominating majority carrier throughout this temperature range \cite{Wang2017}. Nevertheless, the Hall data demonstrates a strong correlation between the magnetic structure and the Fermi surface which leads to the non-monotonic temperature dependence of R$_0$ in F4GT \cite{Li2020c}. 

The temperature-dependent anomalous Hall coefficient (R$_S$), derived from the slope of Eq. \ref{Eq.6} by fitting the Hall data, is shown in Fig.~\ref{fig:Fig3_OK}(d).  R$_S$ shows a non-monotonic behavior with its maximum at $\sim$ 130~K and a strong decrease below the SRT. However, it was argued that the R$_S$ may not describe the proper scaling with M when there is a significant variation of resistivity with T or B and a material-specific scaling factor S$_H$ was introduced where $ R_{S} = S_{H}\rho_{xx}^{2} $  \cite{lee2007r}. Fig.~\ref{fig:Fig3_OK}(d) (red curve) shows the variation of S$_H$ with temperature, which shows weak variation with the temperature above the SRT transition and changes significantly below SRT, indicating its complex dependence with resistivity and magnetism below the SRT.

To investigate the origin of the anomalous Hall effect (AHE), typically one looks at the scaling behavior of $ \rho_{xy}^{AHE} $, $ \rho_{xy}^{AHE} \approx a\rho_{xx}^{\alpha} $\cite{Nagaosa2010a}. Primarily three mechanisms were identified to describe the AHE for ferromagnets or materials with strong spin-orbit coupling (SOC): intrinsic K-L mechanism, extrinsic side-jump mechanism, and extrinsic skew-scattering mechanism \cite{Nagaosa2010a}. Intrinsic K-L mechanism \cite{Karplus1954a} is associated with the anomalous drift due to the finite Berry curvature in the momentum space appearing due to the SOC. It is mostly independent of scattering and solely dependent on the band structure of the crystal~\cite{Nagaosa2010a, Jungwirth2002, Onoda2002}. Extrinsic side-jump mechanism \cite{Berger1970} is related to the deflection of electrons due to scattering from the spin-orbit coupled impurities, and extrinsic skew-scattering mechanism \cite{Smit1955, Smit1958b} is caused by the asymmetric scattering of electrons from the impurities due to the spin-orbit interactions. All of these three mechanisms are associated with the power law $ \rho_{xy}^{AHE} \approx a\rho_{xx}^{\alpha} $. For intrinsic KL mechanism and extrinsic side-jump mechanism $ \alpha  = 2 $ and for extrinsic skew-scattering mechanism $ \alpha  = 1 $ \cite{Nagaosa2010a}.

To get a quantitative estimate, we have calculated the anomalous Hall conductivity ($ |\sigma_{xy}^{AHE}| \approx  \rho_{xy}^{AHE}/\rho_{xx}^{2} \approx \mu_{0}R_{S}M_{S}/\rho_{xx}^{2} $) and the anomalous Hall angle ($ \theta_{AHE} = \sigma_{xy}^{AHE}/\sigma_{xx} $). It must be noted that the anomalous Hall conductivity is overestimated at the higher temperatures (T $>$ T$_{SR}$) as M does not show full saturation even with 9T magnetic field (see Fig. \ref{fig:Fig3_OK}(a)). The temperature dependence of $ |\sigma_{xy}^{AHE}|$ and $ \theta_{AHE}$ exhibit a non-monotonic behavior, having a maximum near T$_{SR}$, as shown in Fig. \ref{fig:Fig4_OK}(b). While at 5 K, $ |\sigma_{xy}^{AHE}|\approx 123~\Omega^{-1}$cm$^{-1} $, it increases and attains a maximum value of $ \sim $ 365.8 $ \Omega^{-1}$cm$^{-1} $ at $\sim$ 80~K, which is sufficiently high among these class of 2D ferromagnetic materials at this higher temperature regime \cite{Kim2018e, Liu2018a}. The low-temperature value is almost three times lower than the expected AHE conductivity in the 'resonant' condition, $\sim e^{2}/ha \approx 390~\Omega^{-1}$cm$^{-1} $ \cite{Onoda2002, Kim2018c} in three dimensions, where e is the electronic charge, h is the Planck constant and a = 9.97 $ \mathring{A} $ \cite{Seo2020b}, being the lattice constant of F4GT. Similarly, at low-temperature, $ \theta_{AH} \sim 0.005\textdegree $ and it becomes maximum ($ \sim0.0308$\textdegree$ $) near $ T_{SR} $ (See Fig. \ref{fig:Fig4_OK}(b)), which is consistent with the previously reported data \cite{Seo2020b}. These results suggest that extrinsic mechanisms might play a significant role in determining the AHE in F4GT. The extrinsic side-jump contribution to the AHE conductivity ($ \sigma_{xy}^{AHE} $) is in the order of $ e^{2}/(ha)*(\varepsilon_{SO}/E_{F}) $ \cite{Nozieres1973}, where $ \varepsilon_{SO} $ is the spin-orbit coupling interaction and $ E_{F} $ is the Fermi energy. For ferromagnetic metals, this ratio $  \varepsilon_{SO}/E_{F} $ is usually small ($ \sim $ 0.01) and so the extrinsic side-jump contribution \cite{Wang2016e, Kim2018c}. It will be interesting to calculate this ratio for this class of van der Waals ferromagnet to get an estimate of the external side-jump contribution. Very recently, DFT-based calculations reveal that all three members of the FGT family exhibit finite Berry curvature in their electronic structure and show significant AHE contributions \cite{Sau2022}. More importantly, F4GT belongs to a different magnetic symmetry class compared to the other two members, having a non-trivial Berry curvature leading to the maximum intrinsic value ($\approx 365.8~\Omega^{-1}$cm$^{-1} $). As the other inelastic scattering contributions almost vanish, the low-temperature AHE conductivity seems to originate from the  intrinsic KL mechanism, albeit with a lower magnitude. In the next, we focus on the observed strong temperature dependence of AHE conductivity.

It is hard to identify the exact mechanism responsible for AHE due to the observed strong temperature dependence of the AHE coefficient below the SRT. The theory of AHE does not include the role of inelastic scattering like electron-electron, electron-magnon, or electron-phonon interaction. As both of our resistivity and MR data indicate a dominant electron-magnon scattering in the intermediate temperature range, the AHE might also have a similar origin. First, $ \rho_{xy}^{AHE} $ vs $ \rho_{xx} $ is plotted to check the scaling behavior of $ \rho_{xy}^{AHE} $ in the temperature range from 5 K to 100 K, as shown in Fig. \ref{fig:Fig4_OK}(a). From the fitting (red curve in Fig. \ref{fig:Fig4_OK}(a)) of the equation $ \rho_{xy}^{AHE} \approx a\rho_{xx}^{\alpha} $, we have determined the value of $ \alpha  = 1.71 $, i, e. almost quadratic dependence of $ \rho_{xy}^{AHE} $ on $ \rho_{xx} $, which indicates that the AHE of F4GT could be originated dominantly from the intrinsic KL mechanism or extrinsic side-jump mechanism, rather than the extrinsic skew-scattering mechanism where $ \rho_{xy}^{AHE} $ linearly dependent on $ \rho_{xx} $. To understand the dominant contributions, we have plotted $ \rho_{xy}^{AHE} $ with $ \rho_{xx} $ (inset of \ref{fig:Fig4_OK}(a)) and fitted with the equation $ \rho_{xy}^{AHE} = a\rho_{xx}+ b\rho_{xx}^2 $, where a is the strength of the skew scattering contribution and b denotes the strength of the side jump/intrinsic contribution. From the fitting, we found a = -0.0181 and b $ \sim $ 604 S cm$^{-1}$, which indicates that the intrinsic Berry phase and/or extrinsic side-jump contribution highly dominates over the skew scattering contribution. Here, the negative sign of $a$ indicates that the skew scattering contribution is acting in the opposite direction as compared to the other two mechanisms.
\begin{figure}[ht!]
	\centering
	\includegraphics[width=8.5cm]{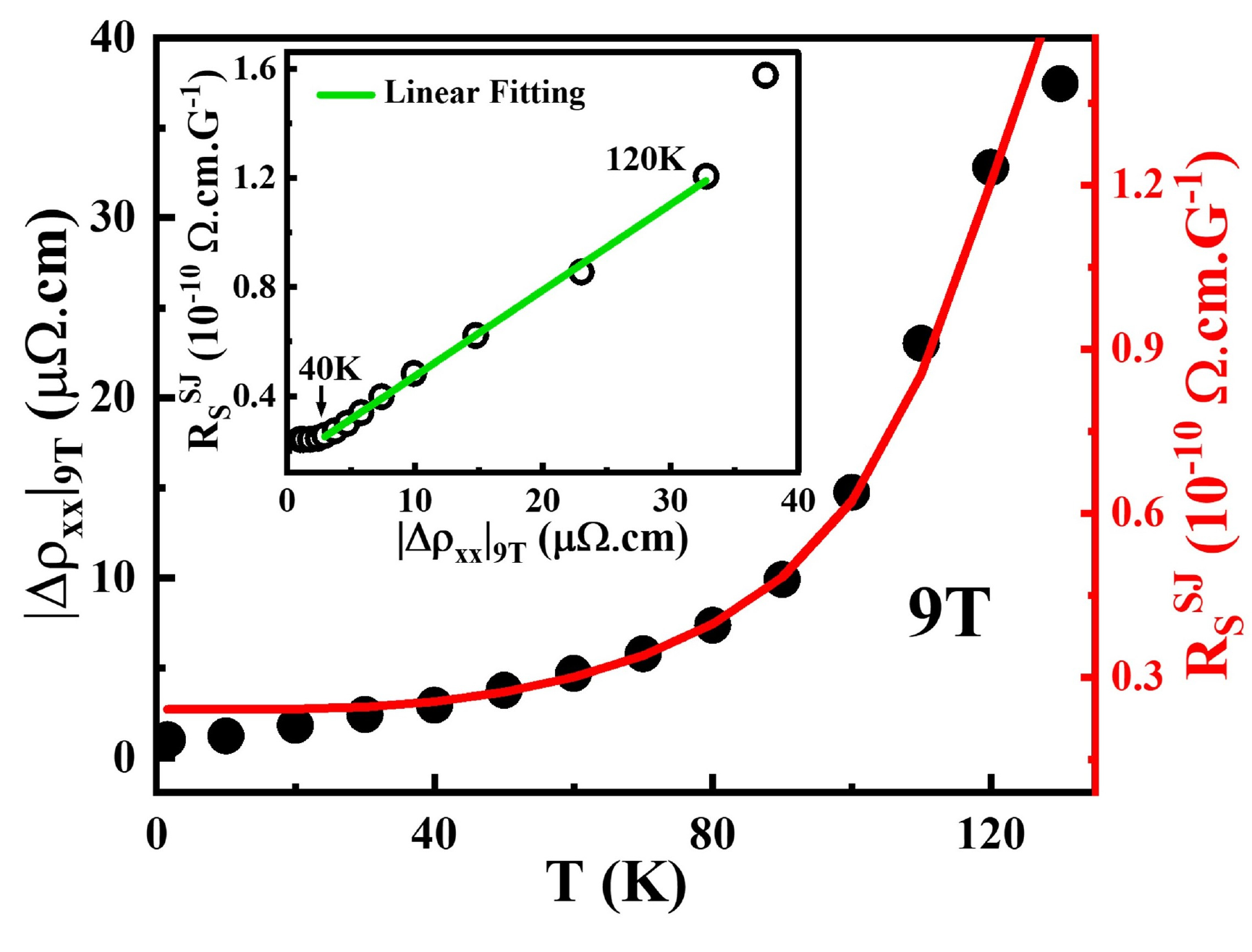}
	\caption{Temperature dependence of the change in longitudinal resistivity ($ |\Delta\rho_{xx}|_{9 T} $) (black solid circles) and extrinsic side-jump contribution ($ R_{S}^{SJ} $) (red line). Inset: The fitting between $ |\Delta\rho_{xx}|_{9 T} $ and $ R_{S}^{SJ} $ shows linear behavior in the temperature range between T$_Q$ and T$_{SR}$.}
	\label{fig:Fig5}
\end{figure}
Since, the extrinsic side-jump contribution is independent of the strength and density of the scatters similar to intrinsic mechanisms and both of them follow the quadratic dependence to the longitudinal resistivity ($ \propto \rho_{xx}^{2} $), it is difficult to separate the intrinsic KL and extrinsic side-jump contributions. However, the intrinsic part typically does not change with temperature in most cases unless there is an electronic transition with temperature leading to non-monotonic Berry phase contribution~\cite{Nagaosa2010a, Miyasato2007}. At this moment, as there is no consensus on the Fermi surface reconstruction and possible electronic transition near SRT, it is not possible to comment on the change in the intrinsic contribution due to SRT. Recently, Yang et. al. \cite{Yang2011a} theoretically proposed that the side-jump contribution can be affected by the electron-magnon scattering, which may lead to the temperature dependence of anomalous Hall conductivity (AHC). To investigate further, we decouple different scattering terms of R$_S$(T) using the procedure mentioned in Ref. \cite{Yang2011a} (for detailed calculation, see Supplementary Information section S6). Fig. \ref{fig:Fig5} shows the temperature dependence plot of change in resistivity ($\Delta \rho_{xx}$) at 9 T magnetic field and R$_S$$^{SJ}$. The red line shows the extracted side jump contribution to the resistivity, which scales perfectly in the temperature range T$_Q$ $<$ T $<$ T$_{SR}$ and deviates in the low-temperature regime (T $<$ T$_Q$) as well as T $>$ T$_{SR}$.  Additionally, R$_S$$^{SJ}$ also shows a linear relation with $ |\Delta\rho_{xx}|_{9 T} $ (Inset in Fig. \ref{fig:Fig5}). The above analysis directly indicates that the R$_S$$^{SJ}$(T) primarily originates from the spin-flip electron–magnon scattering.\\

\section{Conclusion}

In summary, a comprehensive electronic and magneto-transport study is presented in a quasi-two-dimensional van der Waals ferromagnet, F4GT. Temperature-dependent resistivity shows a metal-like behavior with a strong fall after the spin reorientation transition. While at the low-temperature limit (below T$_Q \sim 38$~K), resistivity is governed by pure electron-electron scattering with a quadratic temperature dependence, the inelastic scattering contributions, electron-magnon and electron-phonon scattering, become significant in the intermediate temperature range (T$_Q$ $<$ T $<$ T$_{SR}$). Beyond T$_{SR}$, electron-phonon scattering dominates the resistivity. Similarly, the magnetoresistance data shows distinctive features with positive (T $<$ T$_Q$) to negative (T$_Q$ $<$ T $<$ 300~K) MR having a maximum near the SRT, indicating a dominant orbital contribution at low temperatures and dominant electron-magnon scattering exhibiting in the intermediate temperatures. The temperature-dependent Hall data also displays interesting features, with the ordinary Hall coefficient showing a sharp transition from positive to negative near the SRT and again to positive above T$_Q$, suggesting a significant Fermi surface reconstruction possibly arising due to the SRT. In addition, we observe a large anomalous Hall conductivity with its value ranging from $\approx 123~\Omega^{-1}$cm$^{-1}$ at low temperature ($\sim 5$~K) to a maximum value of $\approx 365~\Omega^{-1}$cm$^{-1} $ near the SRT. While the low-temperature contributions possibly arise due to the intrinsic KL mechanism due to the finite Berry curvature in the electronic structure of F4GT, our analysis reveals the extrinsic side jump mechanism arising due to the spin-flip electron-magnon scattering, is responsible for the strong temperature dependence of AHE in the intermediate temperature range. The detailed transport study not only sheds light on the scattering mechanisms responsible for temperature-dependent behavior resistivity, MR, and Hall effect but also clearly demonstrates a new electronic transition to a pure nonmagnetic metal below $\sim$ T$_Q$  other than the spin re-orientation transition.

\begin{acknowledgments}

This research has made use of “Thematic Unit of Excellence on Nanodevice Technology” (grant no. SR/NM/NS-09/2011) and the Technical Research Centre (TRC) Instrument facilities of S. N. Bose National Centre for Basic Sciences, established under the TRC project of Department of Science and Technology (DST), Govt. of India. The authors acknowledge discussions with Shubhankar Roy, Manoranjan Kumar, Nitesh Kumar, Shubhro Bhattacharjee and Arun Paramekanti.

\end{acknowledgments}

\appendix

\bibliographystyle{apsrev4-2}
\bibliography{manuscript1.bbl}

\end{document}